# Improved Upper Bounds on Stopping Redundancy

Junsheng Han and Paul H. Siegel, *Fellow, IEEE*

*Abstract*—Let $\mathcal{C}$ be a linear code with length $n$ and minimum distance $d$. The *stopping redundancy* of $\mathcal{C}$ is defined as the minimum number of rows in a parity-check matrix for $\mathcal{C}$ such that the smallest stopping sets in the corresponding Tanner graph have size $d$. We derive new upper bounds on the stopping redundancy of linear codes in general, and of maximum distance separable (MDS) codes specifically, and show how they improve upon previously known results. For MDS codes, the new bounds are found by upper bounding the stopping redundancy by a combinatorial quantity closely related to Turán numbers. (The *Turán number*, $T(v,k,t)$, is the smallest number of $t$-subsets of a $v$-set, such that every $k$-subset of the $v$-set contains at least one of the $t$-subsets.) We further show that the stopping redundancy of MDS codes is $T(n, d-1, d-2)(1 + O(n^{-1}))$ for fixed $d$, and is at most $T(n, d-1, d-2)(3 + O(n^{-1}))$ for fixed code dimension $k = n - d + 1$. For $d = 3, 4$, we prove that the stopping redundancy of MDS codes is equal to $T(n, d-1, d-2)$, for which exact formulas are known. For $d = 5$, we show that the stopping redundancy of MDS codes is either $T(n, 4, 3)$ or $T(n, 4, 3) + 1$.

## I. INTRODUCTION

*Stopping sets* play an important role in the performance of iterative decoding of linear codes on erasure channels. Unlike the weight distribution, which is a property of the code itself, stopping sets are dependent on the actual Tanner graph (or parity-check matrix) used to describe the code. In terms of the Tanner graph, a stopping set is a set of variable nodes such that no check node is connected to the set with a single edge. Equivalently, in terms of a parity-check matrix $H$, a stopping set is a set of column indices such that the submatrix formed by the corresponding columns of $H$ does not contain a row of weight one. We will focus on the interpretation in terms of the parity-check matrix. It should be noted that in our context a parity-check matrix can have dependent rows as long as the rows of the matrix span the dual code.

For a chosen $H$, define the size of the smallest non-empty stopping set as its *stopping distance*, denoted by $s(H)$. $s(H)$ is an important parameter for iterative decoding on erasure channels, and should be maximized for better performance. Let $\mathcal{C}$ be a linear code and denote its minimum distance by $d(\mathcal{C})$. Since the support of any codeword is a stopping set, $s(H) \leq d(\mathcal{C})$ for all choices of $H$. We are interested in how many parity checks are needed such that $s(H) = d(\mathcal{C})$ can be achieved. The *stopping redundancy* of the code, denoted by $\rho(\mathcal{C})$, represents the minimum number of rows in a parity-check matrix such that $s(H) = d(\mathcal{C})$. This quantity was defined and first investigated by Schwartz and Vardy [1], [2]. They showed that $\rho(\mathcal{C})$ is well-defined in that by proper choice of $H$, $s(H) = d(\mathcal{C})$ can always be achieved. They then

J. Han and P. H. Siegel are with the University of California, San Diego, La Jolla, CA 92093-0407 (han@cts.ucsd.edu, psiegel@ucsd.edu).

developed general upper and lower bounds on $\rho(\mathcal{C})$, as well as more specific bounds for Reed-Muller codes, Golay codes, and maximum distance separable (MDS) codes. The stopping redundancy of Reed-Muller codes has been further studied by Etzion [3].

In this paper, we propose new upper bounds on $\rho(\mathcal{C})$ and compare them to those in [1]. For the general upper bound, we take a probabilistic approach [4]. In the case of MDS codes, Schwartz and Vardy pointed out a a link between $\rho(\mathcal{C})$ and covering numbers. This led to a number of lower bounds on $\rho(\mathcal{C})$. We show that $\rho(\mathcal{C})$ of MDS codes is upper bounded by another combinatorial quantity. Further results reveal a strong connection between $\rho(\mathcal{C})$ of MDS codes and Turán numbers – combinatorial quantities closely related to covering numbers. Specifically, the *Turán number*, $T(v, k, t)$, is the smallest number of $t$-subsets of a $v$-set, such that every $k$-subset of the $v$-set contains at least one of the $t$-subsets.

In Section II, we derive general upper bounds on $\rho(\mathcal{C})$. First, we consider binary linear codes. We show that an upper bound given in [1] has an interesting variant which is always tighter if $d(\mathcal{C})$ is odd. We then propose another upper bound based on probabilistic analysis. We compare this bound to the other bounds and show that it is better in many interesting cases. In particular, for asymptotically "good" codes, the upper bound derived using the probabilistic method is always tighter. Next, we extend the upper bounds to linear codes over $\mathbb{F}_q$. We prove that, in this context as well, the upper bound based on probabilistic ideas is tighter for "good" codes.

In Section III we focus on MDS codes. First, we recall the observation made in [1] to show that for an MDS code $\mathcal{C}$ with length $n$ and minimum distance $d$, $\rho(\mathcal{C}) \geq T(n, d-1, d-2)$. Next, by introducing a new combinatorial object, we convert the quest for upper bounds on $\rho(\mathcal{C})$ to a purely combinatorial problem. Proceeding in this way, we first discover that the lower bound of $T(n, d-1, d-2)$ is tight for small values of $d$. In particular, for $d = 3, 4$, we prove that $\rho(\mathcal{C}) = T(n, d-1, d-2)$, for which exact formulas are known, and, for $d = 5$, we show that $\rho(\mathcal{C})$ is no greater than $T(n, 4, 3) + 1$. We then generalize these results and show that for a fixed minimum distance $d$, the stopping redundancy of MDS codes is asymptotically equal to $T(n, d-1, d-2)$. Finally, we obtain tighter upper bounds through explicit constructions of the newly defined combinatorial object. One of the upper bounds further shows that for fixed code dimension $k = n - d + 1$, the stopping redundancy of MDS codes is asymptotically at most $T(n, d-1, d-2)(3 + O(n^{-1}))$.

Section IV concludes the paper.

## II. GENERAL BOUNDS

### A. Binary linear codes

Let $r(\mathcal{C})$ denote the *redundancy* of code $\mathcal{C}$, i.e. $r(\mathcal{C}) = \dim(\mathcal{C}^\perp)$, where $\mathcal{C}^\perp$ is the dual code of $\mathcal{C}$. The following theorem is taken from [1].

**Theorem 1** *Let $\mathcal{C}$ be a binary linear code with $d(\mathcal{C}) \geq 3$. Then*

$$\rho(\mathcal{C}) \leq \sum_{i=1}^{d(\mathcal{C})-2} \binom{r(\mathcal{C})}{i}. \tag{1}$$

Following the same idea, we derive the following bound, which is often better than (1).

**Theorem 2** *Let $\mathcal{C}$ be a binary linear code with $d(\mathcal{C}) \geq 2$. Then*

$$\rho(\mathcal{C}) \leq \sum_{i=1}^{\lceil \frac{d(\mathcal{C})-1}{2} \rceil} \binom{r(\mathcal{C})}{2i-1}. \tag{2}$$

*Proof:* Take any basis of $\mathcal{C}^\perp$ to form a parity-check matrix $H$. If $\mathcal{C}$ is of length $n$, then $H$ is an $r(\mathcal{C}) \times n$ matrix. Now, for all odd $i$, $1 \leq i \leq d(\mathcal{C})-1$, take all non-zero linear combinations of $i$ rows of $H$ and put them all together to form a new matrix $H'$. Clearly, $H'$ is a parity-check matrix for $\mathcal{C}$, and the number of rows in $H'$ is exactly the quantity on the right-hand-side of (2).

It suffices to show that $s(H') = d(\mathcal{C})$. For $t = 1, 2, \ldots, d(\mathcal{C})-1$, take an arbitrary set of $t$ columns of $H$ and form the matrix $H_t$. Take the corresponding $t$ columns of $H'$ and form the corresponding matrix $H'_t$. Since $t < d(\mathcal{C})$, the columns of $H_t$ are linearly independent. Therefore, there exist $t$ rows of $H_t$ that form a basis for $\mathbb{F}_2^t$. Take $t$ such rows of $H_t$ and call this $t \times t$ matrix $H_{tt}$. Clearly, $H_{tt}$ is full rank.

By construction, $H'_t$ contains all linear combinations of an odd number of rows of $H_{tt}$. The proof is complete if we can show that at least one of these linear combinations yields a vector of weight one. Think of linearly combining rows of $H_{tt}$ as multiplying $H_{tt}$ by a row vector on the left. Then the linear combinations of rows of $H_{tt}$ that give vectors of weight one are precisely described by the rows of $G$, where $G$ is a $t \times t$ binary matrix such that $GH_{tt} = I$. ($I$ is the $t \times t$ identity matrix.) That is, they are precisely described by the rows of $H_{tt}^{-1}$. The fact that $H_{tt}^{-1}$ is a full rank binary matrix implies that $H_{tt}^{-1}$ must have at least one row of odd weight. ∎

*Remark* If $d(\mathcal{C})$ is odd, then the bound of (2) is always better than (1) as it sums a proper subset of the terms in (1), all of which are positive. If $d(\mathcal{C})$ is even, an improvement is not guaranteed since the bound in (2) includes the term $\binom{r(\mathcal{C})}{d(\mathcal{C})-1}$ while that in (1) does not. For the particular case where $r(\mathcal{C})$ grows with $n$ while $d(\mathcal{C})$ remain fixed, (2) is asymptotically a looser bound. One can, of course, always take the smaller of the two to get best results. ∎

*Remark* Bound (2) implies that $\rho(\mathcal{C}) \leq 2^{r(\mathcal{C})-1}$, an upper bound which can not be deduced from (1). Note that $\rho(\mathcal{C}) \leq 2^{r(\mathcal{C})} - 1$ can be easily shown by considering a parity-check matrix that contains all nonzero codewords of $\mathcal{C}^\perp$. (See [1].) ∎

We now propose another upper bound on $\rho(\mathcal{C})$ based on a probabilistic approach (cf. [4]).

**Theorem 3** *Let $\mathcal{C}$ be a binary linear code with length $n$. Then*

$$\rho(\mathcal{C}) \leq \rho^*(n, d(\mathcal{C})) + r(\mathcal{C}) - d(\mathcal{C}) + 1, \tag{3}$$

*where $\rho^*(n,d)$ is the smallest integer $\rho^*$ that satisfies*

$$\sum_{i=1}^{d-1} \binom{n}{i} \left(1 - \frac{i}{2^i}\right)^{\rho^*} < 1 \tag{4}$$

∎

*Proof:* For any given number of rows, $\rho$, consider a random ensemble of matrices, $\mathcal{H}_\rho$, consisting of all $\rho \times n$ matrices whose rows are codewords of $\mathcal{C}^\perp$. Let the probability measure $P$ on $\mathcal{H}_\rho$ be that which is induced when the rows of matrices in $\mathcal{H}_\rho$ are chosen uniformly and independently from $\mathcal{C}^\perp$.

Let $[n]^i$ denote the set of all $i$-element subsets of $\{1, 2, \ldots, n\}$. Using the terminology of [1], we refer to the elements of $[n]^i$ as *$i$-sets* and think of them as sets of vector coordinates. For a matrix $H$ with $n$ columns, we say that $H$ *covers* $\iota \in [n]^i$ if the projection of rows of $H$ onto $\iota$ contains a vector of weight one. Clearly, $s(H) = d(\mathcal{C})$ if and only if $H$ covers all $i$-sets for $i = 1, \ldots, d(\mathcal{C}) - 1$.

It is well-known [5, p. 139] that the matrix of all codewords of $\mathcal{C}^\perp$ is an orthogonal array of strength $d(\mathcal{C})-1$. This implies that on any $i$-set, $i = 1, \ldots, d(\mathcal{C})-1$, all $i$-tuples appear, and they appear the same number of times. Since there are $i$ weight-one vectors among a total of $2^i$ possible $i$-tuples, the probability that any given $i$-set is covered by a randomly chosen codeword of $\mathcal{C}^\perp$ is $i/2^i$. Hence, for $i = 1, \ldots, d(\mathcal{C})-1$, the probability that a given $i$-set is not covered by rows in a matrix in the random ensemble $\mathcal{H}_\rho$ is $(1 - i/2^i)^\rho$. We have

$$P(\{\text{all } i\text{-sets are covered}, i = 1, \ldots, d(\mathcal{C})-1\}) \tag{5}$$

$$= 1 - P(\{\text{at least one } i\text{-set is not covered}$$
$$\text{for some } i \in \{1, \ldots, d(\mathcal{C})-1\}\}) \tag{6}$$

$$= 1 - P\left(\bigcup_{i=1}^{d(\mathcal{C})-1} \bigcup_{\iota \in [n]^i} \{\iota \text{ is not covered}\}\right) \tag{7}$$

$$\geq 1 - \sum_{i=1}^{d(\mathcal{C})-1} \sum_{\iota \in [n]^i} \left(1 - \frac{i}{2^i}\right)^\rho \tag{8}$$

$$= 1 - \sum_{i=1}^{d(\mathcal{C})-1} \binom{n}{i} \left(1 - \frac{i}{2^i}\right)^\rho. \tag{9}$$

If $\sum_{i=1}^{d(\mathcal{C})-1} \binom{n}{i}(1 - \frac{i}{2^i})^\rho < 1$, then $P(\{\text{all } i\text{-sets are covered}, i = 1, \ldots, d(\mathcal{C})-1\}) > 0$, which implies that there exists $H \in \mathcal{H}_\rho$ that covers all $i$-sets, $i = 1, \ldots, d(\mathcal{C})-1$. Note that the fact that $H$ covers all $i$-sets up to $i = d(\mathcal{C})-1$ implies that $\text{rank}(H) \geq d(\mathcal{C})-1$. Therefore, by adding at most $r(\mathcal{C}) - d(\mathcal{C}) + 1$ appropriate codewords from $\mathcal{C}^\perp$ as additional rows to $H$, we have found a parity-check matrix for $\mathcal{C}$ that covers all $i$-sets, $i = 1, \ldots, d(\mathcal{C}) - 1$. ∎

The upper bound given in Theorem 3 involves solving an inequality. A closed form expression would be desirable. This is addressed in the following corollaries.

**Corollary 4** *Let $\mathcal{C}$ be a binary linear code with length $n$ and minimum distance $d(\mathcal{C}) < n/2$. Then*

$$\rho(\mathcal{C}) \leq \frac{nh(\delta) + \frac{1}{2}\log\frac{\delta}{2\pi n(1-\delta)(1-2\delta)^2}}{-\log\left(1 - \frac{d(\mathcal{C})-1}{2^{d(\mathcal{C})-1}}\right)} + r(\mathcal{C}) - d(\mathcal{C}) + 1, \quad (10)$$

*where $\delta = \frac{d(\mathcal{C})}{n}$, and $h(\delta) = \delta \log \frac{1}{\delta} + (1-\delta)\log\frac{1}{1-\delta}$.* □

*Proof:* First note that $(1 - i/2^i)$ is non-decreasing in $i$, so that

$$\sum_{i=1}^{d-1}\binom{n}{i}\left(1 - \frac{i}{2^i}\right)^\rho \leq \left(1 - \frac{d-1}{2^{d-1}}\right)^\rho \sum_{i=1}^{d-1}\binom{n}{i}. \quad (11)$$

Next, for $0 < \delta = \frac{d(\mathcal{C})}{n} < \frac{1}{2}$, it can be shown that

$$\sum_{i=1}^{d(\mathcal{C})-1}\binom{n}{i} < \frac{\delta}{1-2\delta}\binom{n}{\delta n}. \quad (12)$$

Further, by Stirling's approximation it is known that ([6])

$$\binom{n}{\delta n} \leq \frac{1}{\sqrt{2\pi n\delta(1-\delta)}} 2^{nh(\delta)}. \quad (13)$$

Now, by putting together (11), (12), and (13), and referring to (4), we see that a positive solution to the equation

$$\frac{\delta}{1-2\delta}\frac{1}{\sqrt{2\pi n\delta(1-\delta)}} 2^{nh(\delta)}\left(1 - \frac{d(\mathcal{C})-1}{2^{d(\mathcal{C})-1}}\right)^\rho = 1. \quad (14)$$

must be an upper bound on $\rho^*(n, d(\mathcal{C}))$. We thus obtain

$$\rho^*(n, d(\mathcal{C})) \leq \frac{nh(\delta) + \frac{1}{2}\log\frac{\delta}{2\pi n(1-\delta)(1-2\delta)^2}}{-\log\left(1 - \frac{d(\mathcal{C})-1}{2^{d(\mathcal{C})-1}}\right)}. \quad (15)$$

Plugging (15) in (3) we get the desired bound. ■

If we do not require $d(\mathcal{C}) < n/2$, we have to weaken the upper bound, but the resulting bound has a simpler form.

**Corollary 5** *Let $\mathcal{C}$ be a binary linear code with length $n$. Then*

$$\rho(\mathcal{C}) \leq \frac{n}{-\log\left(1 - \frac{d(\mathcal{C})-1}{2^{d(\mathcal{C})-1}}\right)} + r(\mathcal{C}) - d(\mathcal{C}) + 1. \quad (16)$$

□

*Proof:* The argument is almost identical to the proof of Corollary 4, except that we instead bound $\sum_{i=1}^{d(\mathcal{C})-1}\binom{n}{i}$ by

$$\sum_{i=1}^{d(\mathcal{C})-1}\binom{n}{i} < 2^n. \quad (17)$$

■

*Remark* While the bounds in Theorem 1 and Theorem 2 are roughly on the same order, the upper bound in Theorem 3 often appears to be tighter than both. We demonstrate this for a specific example — the extended binary Golay code — and for two asymptotic scenarios ($d(\mathcal{C})$ and $r(\mathcal{C})$ both linear in $n$, and $d(\mathcal{C})$ fixed).

**Example 1** Let $\mathcal{G}_{24}$ denote the extended binary Golay $(24, 12, 8)$ code. In [1], it was shown by explicit construction that $\rho(\mathcal{G}_{24}) \leq 35$. This was later improved to $\rho(\mathcal{G}_{24}) \leq 34$ [2].

Applying the upper bounds obtained in this section to $\mathcal{G}_{24}$, we see that Theorem 1 gives $\rho(\mathcal{G}_{24}) \leq 2509$, Theorem 2 gives $\rho(\mathcal{G}_{24}) \leq 1816$, and Theorem 3 gives $\rho(\mathcal{G}_{24}) \leq 232$. Also, the relaxed bounds in Corollary 4 and Corollary 5 give $\rho(\mathcal{G}_{24}) \leq 245$ and $\rho(\mathcal{G}_{24}) \leq 300$, respectively. We see that in this example, bounds based on Theorem 3 have a clear advantage. □

*Remark* A 34-row parity-check matrix for $\mathcal{G}_{24}$ that achieves maximum stopping distance is given in Appendix I. Compared to the one reported in [2], this parity-check matrix is able to correct more low-weight erasure patterns. □

**Example 2** The bound of Theorem 3 is a function of $n$, $d(\mathcal{C})$ and $r(\mathcal{C})$. Similarly, the bounds of Theorem 1 and Theorem 2 are functions of $d(\mathcal{C})$ and $r(\mathcal{C})$. In this example we consider the asymptotic behavior of these bounds as $n \to \infty$. Detailed derivations can be found in Appendix II.

We discuss two different assumptions about $d(\mathcal{C})$ and $r(\mathcal{C})$. The first case corresponds to "good" codes, i.e. codes whose rate is bounded away from zero and whose minimum distance is non-diminishing relative to the code length. The second case concerns codes with fixed minimum distance, an example of which is the family of extended binary Hamming codes.

**Case 1:** $d(\mathcal{C}) = \delta n$, $r(\mathcal{C}) = \gamma n$, where $0 < \delta < \frac{1}{2}$, $0 < \gamma < 1$ *are constants.*

It can be shown that the bound in Theorem 3 is $\Theta(2^{\delta n})$. In comparison, the bounds of Theorem 1 and Theorem 2 are both $\Omega\left(\frac{2^{2\delta n}}{\sqrt{n}}\right)$. Clearly, the bound given by Theorem 3 is tighter.

**Case 2:** $d(\mathcal{C}) = d$ *is a constant.*

With the expression in Corollary 4, it is not hard to see that the bound of Theorem 3 is $\Theta(\log n + r(\mathcal{C}))$. On the other hand, the bound given by Theorem 1 is clearly $\Theta(r(\mathcal{C})^{d-2})$; and the bound given by Theorem 2 is $\Theta(r(\mathcal{C})^{d-2})$ if $d$ is odd, and $\Theta(r(\mathcal{C})^{d-1})$ if $d$ is even.

By the Hamming bound, $r(\mathcal{C}) > \log n$ for $d \geq 3$. Therefore, as long as $d > 3$, the bound by Theorem 3 is asymptotically tighter. Since it is known for all binary linear codes [1] that if $d(\mathcal{C}) \leq 3$, then $\rho(\mathcal{C}) = r(\mathcal{C})$, Theorem 3 gives a better bound asymptotically for all non-trivial values of $d$. □

### B. Linear codes over $\mathbb{F}_q$

The bounds in Theorem 1 and Theorem 2 can both be viewed as improved versions of the more intuitive bound $\rho(\mathcal{C}) \leq \sum_{i=1}^{d(\mathcal{C})-1}\binom{r(\mathcal{C})}{i}$, which extends in a straightforward manner to non-binary codes (although, unfortunately, the improvements made in Theorem 1 and Theorem 2 cannot be carried over).

**Theorem 6** *Let $\mathcal{C}$ be a linear code over $\mathbb{F}_q$. Then*

$$\rho(\mathcal{C}) \leq \sum_{i=1}^{d(\mathcal{C})-1}\binom{r(\mathcal{C})}{i}(q-1)^{i-1}. \quad (18)$$

□

*Proof:* The proof is similar to that of Theorem 2. Here we take a basis of $\mathcal{C}^\perp$ and construct $H$ by taking linear combinations of $i$ basis vectors, for $i = 1, \ldots, d(\mathcal{C})-1$. Instead of taking *all* linear combinations, note that for each $i$ basis vectors we may fix one of the linear coefficients at 1 and still be guaranteed that $s(H) = d(\mathcal{C})$. ∎

For $\mathcal{C}$ a linear code over $\mathbb{F}_q$, the codewords of $\mathcal{C}^\perp$ are known to form an orthogonal array of strength $d(\mathcal{C}) - 1$ with $q$ levels ([7, ch. 4]). Therefore, the argument we used to prove Theorem 3 extends directly to non-binary codes.

**Theorem 7** *Let $\mathcal{C}$ be a linear code over $\mathbb{F}_q$ with length $n$. Then*

$$\rho(\mathcal{C}) \leq \rho^*(n, d(\mathcal{C}), q) + r(\mathcal{C}) - d(\mathcal{C}) + 1, \quad (19)$$

*where $\rho^*(n, d, q)$ is the smallest integer $\rho^*$ that satisfies*

$$\sum_{i=1}^{d-1} \binom{n}{i} \left(1 - \frac{(q-1)i}{q^i}\right)^{\rho^*} < 1. \quad (20)$$

∎

**Corollary 8** *Let $\mathcal{C}$ be a linear code over $\mathbb{F}_q$ with length $n$ and minimum distance $d(\mathcal{C}) < n/2$. Then*

$$\rho(\mathcal{C}) \leq \frac{nh(\delta) + \frac{1}{2}\log\frac{\delta}{2\pi n(1-\delta)(1-2\delta)^2}}{-\log\left(1 - \frac{(q-1)(d(\mathcal{C})-1)}{q^{d(\mathcal{C})-1}}\right)} + r(\mathcal{C}) - d(\mathcal{C}) + 1, \quad (21)$$

*where $\delta = \frac{d(\mathcal{C})}{n}$, and $h(\delta) = \delta\log\frac{1}{\delta} + (1-\delta)\log\frac{1}{1-\delta}$.* ∎

**Corollary 9** *Let $\mathcal{C}$ be a linear code over $\mathbb{F}_q$ with length $n$. Then*

$$\rho(\mathcal{C}) \leq \frac{n}{-\log\left(1 - \frac{(q-1)(d(\mathcal{C})-1)}{q^{d(\mathcal{C})-1}}\right)} + r(\mathcal{C}) - d(\mathcal{C}) + 1. \quad (22)$$

∎

**Example 3** Let $\mathcal{G}_{12}$ denote the ternary (12,6,6) Golay code. The bound of Theorem 6 gives $\rho(\mathcal{G}_{12}) \leq 332$, while the bound of Theorem 7 gives $\rho(\mathcal{G}_{12}) \leq 160$. The best known result (by construction, see [1]) is $\rho(\mathcal{G}_{12}) \leq 22$. ∎

**Example 4** Similar to Example 2 for the case of binary codes, we compare the bounds of Theorem 6 and Theorem 7 as $n \to \infty$. Here we will only treat the case of "good" codes.

Let $d(\mathcal{C}) = \delta n$, $r(\mathcal{C}) = \gamma n$, where $0 < \delta < \frac{q-1}{q}$ and $0 < \gamma < 1$ are constants. It is not hard to show that the bound of Theorem 7 is $\Theta(q^{\delta n})$. On the other hand, it can be shown (details provided in Appendix II) that the bound of Theorem 6 is $\Omega\left(\frac{1}{\sqrt{n}} q^{\frac{q}{q-1}\delta n}\right)$. We see that the bound given by Theorem 7 is tighter. ∎

## III. MDS CODES

Being MDS imposes a lot of structure on a code. We will take advantage of the special properties of MDS codes to show that their stopping redundancy is of a highly combinatorial nature and is closely related to Turán numbers. New, tighter upper bounds will be obtained through constructions.

First a few notes (reminders) on notation. Let $n$, $k$ be integers and $A$, $B$ be sets. Then
- $|A| :=$ Number of elements of $A$.
- $A \setminus B := \{x \in A : x \notin B\}$.
- $[n] := \{1, 2, \ldots, n\}$.
- $[A]^k := \{X \subseteq A : |X| = k\}$ is the set of $k$-subsets of $A$.
- $[n]^k := [[n]]^k$.

Also, a $k$-*set* is generally any set that has $k$ elements. Particular to our discussions, a $k$-set usually refers to a set of $k$ codeword coordinates, i.e. a $k$-subset of $[n]$, if $n$ is the length of the code.

A *Turán* $(v, k, t)$-*system* is a set of $t$-subsets of a $v$-set, called *blocks*, such that each $k$-subset of the $v$-set contains at least one of the blocks. The smallest number of blocks in a Turán $(v, k, t)$-system is known as the *Turán number*, and is correspondingly denoted by $T(v, k, t)$. For more information on Turán numbers, the reader is referred to [8], and references therein.

Consider an MDS code $\mathcal{C}$ of length-$n$ and minimum distance $d$. Then its dual code, $\mathcal{C}^\perp$, is an MDS code with minimum distance $d^\perp = n - d + 2$. Also, note that for all MDS codes with minimum distance $d$, any set of $d$ coordinates is the support of at least one codeword. These properties (and many more) can be found in MacWilliams and Sloane [5].

The authors of [1] noted the following.[1]

**Theorem 10** *Let $\mathcal{C}$ be a MDS code with length $n$ and minimum distance $d$. Then*

$$\rho(\mathcal{C}) \geq T(n, d-1, d-2). \quad (23)$$

∎

*Proof:* Suppose $H$ is a parity-check matrix for $\mathcal{C}$ and $s(H) = d$. Then each row of $H$ has at most $n - d^\perp = d - 2$ zeros. If $h$ is a row of $H$ and $\iota$ is a $(d-1)$-set, then $h$ covers $\iota$ if and only if $h$ has $(d-2)$ zeros and the positions of the zeros in $h$ are a subset of $\iota$. Since all $(d-1)$-sets are covered by $H$, the complements of supports of minimum-weight rows of $H$ form a Turán $(n, d-1, d-2)$-system. ∎

This link between stopping redundancy and Turán numbers immediately gives rise to a number of lower bounds on $\rho(\mathcal{C})$ for MDS codes. For example, it is simple to note $T(v, k, t) \geq \binom{v}{k}/\binom{v-t}{k-t} = \binom{v}{t}/\binom{k}{t}$. So we immediately obtain

$$\rho(\mathcal{C}) \geq T(n, d-1, d-2) \geq \frac{1}{d-1}\binom{n}{d-2} \quad (24)$$

(cf. [1]). Better bounds can be obtained by utilizing a stronger lower bound on $T(v, k, t)$.

Now, let $\mathcal{C}$ be an MDS code with length $n$ and minimum distance $d$, and consider the minimum number of rows in a parity-check matrix for $\mathcal{C}$ all of whose rows are minimum-weight codewords of $\mathcal{C}^\perp$ and that achieves the maximum stopping distance $d$. This number only depends on $n$ and $d$, because

1) as far as covering $i$-sets is concerned, only the *supports* of rows of the parity-check matrix matter;
2) for any $d^\perp$-set as support, we can find at least one codeword in $\mathcal{C}^\perp$;
3) all rows of a parity-check matrix that achieves $\rho(\mathcal{C})$ must have distinct supports.

---
[1]In [1], the observation was made with respect to covering numbers rather than Turán numbers. A $(v, k, t)$ *covering design* is a set of $k$-subsets of a $v$-set, such that each $t$-subset of the $v$-set is contained in at least one of the $k$-subsets. The smallest size of a covering design is known as the *covering number*, and is correspondingly denoted by $C(v, k, t)$. It is simple to note that a $(v, k, t)$ covering design is a Turán $(v, v-t, v-k)$-system and vice versa. Hence, $C(v, k, t) = T(v, v-t, v-k)$. For more information on covering designs and covering numbers, the reader is referred to [9].

Let us denote this number by $\Gamma'(n,d)$. Clearly, $\Gamma'(n,d)$ is an upper bound of $\rho(\mathcal{C})$. Note that $\Gamma'(n,d)$ always exists since a matrix consisting of one codeword from $\mathcal{C}^\perp$ for each $d^\perp$-set as support achieves stopping distance equal to $d$ (cf. [1]).

We shall see that $\Gamma'(n,d)$ is in fact a combinatorial quantity with a formulation similar to that of Turán numbers, without any explicit reference to codes at all.

**Definition 1** A *single-exclusion* $(v,r)$-*system* is a collection of $r$-subsets of a $v$-set, called *blocks*, such that for all $i$, $i = 1, \ldots, r+1$, each $i$-subset of the $v$-set is covered by at least one of the blocks. Here, an $i$-subset $\iota$ is *covered* by block $\beta$ if
$$|\iota \setminus \beta| = 1. \qquad (25)$$
The smallest number of blocks in a single-exclusion $(v,r)$-system is called the *single-exclusion number*, and is denoted by $\Gamma(v,r)$. □

*Remark* Clearly, condition (25) is equivalent to
$$|\iota \cap \beta| = i - 1. \qquad (26)$$
□

*Remark* The definition of single-exclusion $(v,r)$-system requires that $r \leq v-1$. For $r = v-1$, it is easy to see that $\Gamma(v, v-1) = v$. For the sake of discussion, unless otherwise noted, we shall always make the assumption that $r \leq v-2$. In relation to $\rho(\mathcal{C})$, we are mostly interested in $\Gamma(n, d-2)$, where $n$ is the length of $\mathcal{C}$ and $d$ is the minimum distance. Clearly, $d - 2 \leq n - 2$ is always satisfied. □

*Remark* A single-exclusion $(v,r)$-system is always a Turán $(v, r+1, r)$-system. It is interesting that the definition of single-exclusion systems may actually be interpreted meaningfully in design theory terms. One can analogously define $k$-exclusion $(v,r)$-systems. □

Let $H$ be a parity-check matrix for $\mathcal{C}$ that achieves stopping distance $d$ and whose rows all have weight $d^\perp$. Then the positions of zeros in the rows of $H$ form a single-exclusion $(n, d-2)$-system. On the other hand, let $S$ be a single-exclusion $(n, d-2)$-system. For each $\beta \in S$, we can find $c \in \mathcal{C}^\perp$ such that the support of $c$ is $[n] \setminus \beta$. If we use these codewords as rows to form matrix $H$, then $s(H) = d$. Note that $s(H) = d$ implies that $H$ has a $(d-1) \times (d-1)$ upper triangular submatrix (up to column permutations) and hence $\text{rank}(H) \geq d - 1 = r(\mathcal{C}^\perp)$. Therefore, $H$ is indeed a parity-check matrix. In summary, an $l$-block single-exclusion $(n, d-2)$-system exists if and only if an $l$-row parity-check matrix consisting solely of minimum weight codewords of $\mathcal{C}^\perp$ can be found that achieves maximum stopping distance. Relating to the earlier definition, it is clear that $\Gamma'(n,d) = \Gamma(n, d-2)$.

The following comes straight from the discussions above.

**Theorem 11** *if $\mathcal{C}$ is an MDS code with length $n$ and minimum distance $d$, then*
$$\rho(\mathcal{C}) \leq \Gamma(n, d-2). \qquad (27)$$
□

We conjecture that equality holds always.

**Conjecture 12** *If $\mathcal{C}$ is an MDS code with length $n$ and minimum distance $d$, then*
$$\rho(\mathcal{C}) = \Gamma(n, d-2). \qquad (28)$$
□

Up to now we have bounded $\rho(\mathcal{C})$ between two well-defined combinatorial quantities, $T(n, d-1, d-2)$ and $\Gamma(n, d-2)$. Clearly, any lower bound on $T(n, d-1, d-2)$ is a lower bound on $\rho(\mathcal{C})$ and any upper bound on $\Gamma(n, d-2)$ is an upper bound on $\rho(\mathcal{C})$. We will actually proceed in this way – in fact we will be focusing solely on the upper bound, and all results we shall show for $\rho(\mathcal{C})$ hold for $\Gamma(n, d-2)$ as well, although it may not be made explicit.

We start by looking at how things work for $d = 3, 4, 5$, where much stronger results can be derived.

The case where $d = 3$ is quite trivial, and the result is actually implied by the best upper and lower bounds on $\rho(\mathcal{C})$ given in [1].

**Theorem 13** *Let $\mathcal{C}$ be an MDS code with length $n$ and minimum distance $d = 3$. Then*
$$\rho(\mathcal{C}) = T(n, 2, 1) = n - 1. \qquad (29)$$
□

*Proof:* It suffices to show that $n - 1 \leq T(n, 2, 1) \leq \Gamma(n, 1) \leq n - 1$. On one hand, it is easy to verify that any $(n-1)$-subset of $[n]^1$ is a single-exclusion $(n, 1)$-system. On the other hand, a Turán $(n, 2, 1)$-system cannot have $(n-2)$ or fewer blocks, or there would exist $i, j \in [n]$, such that $\{i, j\}$ does not contain any of the blocks. ■

The case for $d = 4$ needs a bit more work.

**Lemma 14** *For all $n \geq 3$*,
$$T(n, 3, 2) \leq \binom{n-3}{2} + 3. \qquad (30)$$
□

*Proof:* The proof is by construction. Let $L = \{1, 2, 3\}$, $R = [n] \setminus L$, and $T = [L]^2 \cup [R]^2$. It is easy to verify that $T$ is a Turán $(n, 3, 2)$-system, and it has $\binom{n-3}{2} + 3$ blocks. ■

**Theorem 15** *Let $\mathcal{C}$ be a MDS code with length $n \geq 6$ and minimum distance $d = 4$. Then*
$$\rho(\mathcal{C}) = T(n, 3, 2) = \left\lfloor \frac{n}{2} \right\rfloor \left( \left\lceil \frac{n}{2} \right\rceil - 1 \right). \qquad (31)$$
□

*Proof:* The formula for $T(n, 3, 2)$ is a known result first discovered by Mantel [10] in 1907. Later, Turán [11] [12] solved the more general case of $T(n, k, 2)$.

It suffices to show that $\Gamma(n, 2) \leq T(n, 3, 2)$. Let $T$ be a Turán $(n, 3, 2)$-system with smallest size. We show that $T$ must also be a single-exclusion $(n, 2)$-system. By definition of $T$, all 3-sets are covered. We show that all 1- and 2-sets are covered as well.

Suppose there is a 1-set, say $\{i\}$, that is not covered. Then $i$ is contained in all blocks of $T$. But this implies that all 3-subsets of $[n] \setminus \{i\}$ are not covered, contradicting the fact that $T$ is a Turán $(n, 3, 2)$-system.

Suppose there is a 2-set, say $\{i,j\}$, that is not covered. This implies that a block of $T$ either is $\{i,j\}$, or is disjoint from $\{i,j\}$. Note that $\{i,j\}$ must be a block of $T$, or 3-sets like $\{i,j,k\}$ would not be covered. Also, all 2-sets disjoint from $\{i,j\}$ must be blocks of $T$; otherwise, if $\{k,l\} \subseteq [n] \setminus \{i,j\}$ is not a block, then 3-set $\{i,k,l\}$ would not be covered by $T$. This shows that $T(n,3,2) = |T| = \binom{n-2}{2}+1$. But $\binom{n-2}{2}+1 > \binom{n-3}{2} + 3$ for $n \geq 6$, which contradicts Lemma 14. ∎

*Remark* Since the formula for $T(n,3,2)$ is known, Lemma 14 may seem unnecessary. But we find its simple construction to be appealing, and the bound it gives, though loose, is enough to show $\Gamma(n,2) = T(n,3,2)$ without further knowledge about $T(n,3,2)$. □

*Remark* The proof of Theorem 15 needs $n \geq 6$ to go through. It turns out that the only two cases for $n < 6$ are indeed "anomalies" for which $\rho(\mathcal{C})$ is strictly greater than $T(n,3,2)$.

For $n = 4$, $T(4,3,2) = 2$, while it is simple to see that $\rho(\mathcal{C}) = 3$. For $n = 5$, $T(5,3,2) = 4$. But it can be shown that $\rho(\mathcal{C}) = 5$. □

For $d = 5$, we first note a couple of bounds on $T(n,4,3)$.

**Lemma 16**

$$T(n,4,3) \leq \left\lfloor \frac{n}{3} \right\rfloor \left\lfloor \frac{n-1}{3} \right\rfloor \left( 2 \left\lfloor \frac{n-2}{3} \right\rfloor + 1 \right), \quad (32)$$

*where equality holds for $n \leq 13$.* □

*Proof:* The upper bound comes from a construction of Turán $(n,4,3)$-systems due to Ringel [13], which has been verified to be optimal for $n \leq 13$ ([9]). ∎

**Lemma 17** *For $n \geq 13$,*

$$T(n,4,3) \geq \frac{56}{143} \binom{n}{3}. \quad (33)$$

□

*Proof:* It is known ([14]) that $T(n,k,r)/\binom{n}{r}$ is non-decreasing in $n$, hence

$$T(n,k,r) \geq \frac{T(n_0,k,r)}{\binom{n_0}{r}} \binom{n}{r}, \quad \text{for } n \geq n_0. \quad (34)$$

Since $T(13,4,3) = 112$ by Lemma 16, the result follows. ∎

**Theorem 18** *Let $\mathcal{C}$ be an MDS code with length $n$ and minimum distance $d = 5$. Then*

$$T(n,4,3) \leq \rho(\mathcal{C}) \leq T(n,4,3) + 1. \quad (35)$$

*Further,*

$$\rho(\mathcal{C}) = T(n,4,3), \quad \text{for } n = 6, \ldots, 53. \quad (36)$$

□

*Proof:* It suffices to show that $\Gamma(n,3) \leq T(n,4,3)+1$, and $\Gamma(n,3) = T(n,4,3)$ for $n = 6, \ldots, 53$.

For $n = 5$, it is known that $T(5,4,3) = 3$, while it can be easily verified that $\Gamma(5,3) = 4$. So the claimed inequality holds for $n = 5$.

In the following, assume $n \geq 6$. Let $T$ be a Turán $(n,4,3)$-system of smallest size. If $T$ is a single-exclusion $(n,3)$-system then we are done. Otherwise, let $\iota$ be a *smallest i-set* that is not covered. Then $|\iota| = 1$, 2, or 3. (All 4-sets are covered since $T$ is a Turán $(n,4,3)$-system.)

First, suppose $|\iota| = 1$. Since $\iota$ is not covered, it is contained in all blocks of $T$. Then a 4-subset of $[n] \setminus \iota$ is not covered. This is a contradiction.

Next, suppose $|\iota| = 2$, say $\iota = \{i,j\}$. Then any block of $T$ either contains $\iota$ or is disjoint from $\iota$. Out of the $(n-2)$ 3-sets that contain $\iota$, at least $(n-3)$ must be in $T$. Otherwise we could find $a, b \in [n] \setminus \iota$ such that $\{i,j,a\}, \{i,j,b\} \notin T$. But then the 4-set $\{i,j,a,b\}$ would not be covered. On the other hand, all of the $\binom{n-2}{3}$ 3-sets that are disjoint from $\iota$ must be blocks of $T$. Otherwise, if $\{a,b,c\} \subseteq [n] \setminus \iota$ is not a block, then $\{i,a,b,c\}$ would not be covered. In summary, $T$ must have at least $\binom{n-2}{3} + n - 3$ blocks. Since $\binom{n-2}{3} + n - 3 > \lfloor \frac{n}{3} \rfloor \lfloor \frac{n-1}{3} \rfloor \left(2 \lfloor \frac{n-2}{3} \rfloor + 1\right)$ for $n \geq 6$, this contradicts Lemma 16.

Lastly, suppose $|\iota| = 3$, say $\iota = \{i,j,k\}$. Then for all $\beta \in T$, $|\beta \cap \iota| \neq 2$. Note the following facts:

Fact 1: $\iota$ itself must be a block of $T$, otherwise 4-sets like $\{i,j,k,a\}$ would not be covered.

Fact 2: For each 2-set $\{a,b\} \subseteq [n] \setminus \iota$, at least two of $\{a,b,i\}$, $\{a,b,j\}$, and $\{a,b,k\}$ must be blocks of $T$. This is true because if, say, $\{a,b,i\}$ and $\{a,b,j\}$ both were not blocks of $T$, then $\{a,b,i,j\}$ would not be covered.

Fact 3: All blocks that are disjoint from $\iota$ form a Turán $(n-3,4,3)$-system.

Together, these imply that $T(n,4,3) = |T| \geq 1 + 2\binom{n-3}{2} + T(n-3,4,3)$, which contradicts Lemma 16 and Lemma 17 for $n = 6, \ldots, 53$.

For $n \geq 54$, we do not have an immediate contradiction. However, note that a 3-set that contains zero or one element of $\iota$ is covered due to Fact 2, and one that contains two elements of $\iota$ is covered due to Fact 1. So, in this case $\iota$ must be the only 3-set that is not covered. Since $\iota$ is also the smallest uncovered i-set, by adding one more block to $T$ to cover $\iota$, we have found a single-exclusion $(n,3)$-system that has $T(n,4,3)+1$ blocks. ∎

**Corollary 19** *Let $\mathcal{C}$ be an MDS code with length $n$ and minimum distance $d = 5$. Then*

$$\rho(\mathcal{C}) = \left\lfloor \frac{n}{3} \right\rfloor \left\lfloor \frac{n-1}{3} \right\rfloor \left(2 \left\lfloor \frac{n-2}{3} \right\rfloor + 1\right), \quad \text{for } n = 6, \ldots, 13. \quad (37)$$

□

We have seen that $\Gamma(n, d-2)$ (and hence $\rho(\mathcal{C})$ of an MDS code with the corresponding parameters) is almost the same as $T(n, d-1, d-2)$ for small values of $d$. We now show that these results can be generalized in an asymptotic sense when $d$ is fixed.

**Theorem 20** *For fixed $d$, as $n \to \infty$,*

$$\Gamma(n, d-2) = T(n, d-1, d-2)(1 + O(n^{-1})). \quad (38)$$

□

*Proof:* We show that we can always add $O(n^{d-3})$ blocks to a Turán $(n, d-1, d-2)$-system to make it a single-exclusion $(n, d-2)$-system.

Let $L = \{1, \ldots, d-2\}$, and $R = [n] \setminus L$. Let $T' = \{\beta \in [n]^{d-2} : \beta \cap L \neq \emptyset\}$. Clearly,

$$|T'| = \sum_{m=0}^{d-3} \binom{d-2}{d-2-m}\binom{n-d+2}{m} = O(n^{d-3}). \quad (39)$$

We show that blocks of $T'$ cover all $i$-sets, $i = 1, 2, \ldots, d-2$. Let $\iota$ be an $i$-set and $a \in \iota$ be an arbitary element. Take $\iota \setminus \{a\}$, adjoin to it the $(d-i-1)$ smallest elements of $[n] \setminus \iota$ and call the resulting set $\beta$. It is easy to verify that $\beta \in T'$ and $|\iota \setminus \beta| = 1$.

Now, let $T$ be a Turán $(n, d-1, d-2)$-system of smallest size. Let $S = T \cup T'$. Then $S$ is a single-exclusion $(n, d-2)$-system with $T(n, d-1, d-2) + O(n^{d-3})$ blocks.

Finally, note that $T(n, d-1, d-2) = \Theta(n^{d-2})$, since

$$\frac{1}{d-1}\binom{n}{d-2} \leq T(n, d-1, d-2) \leq \binom{n}{d-2}, \quad (40)$$

and the result follows. ∎

With Theorem 20 the following result is immediate.

**Theorem 21** *Let $\{\mathcal{C}_i\}_{i=1}^{\infty}$ be a sequence of MDS codes with strictly increasing code length $\{n_i\}_{i=1}^{\infty}$. If $d(\mathcal{C}_i) = d$ for all $i$, then as $i \to \infty$,*

$$\rho(\mathcal{C}_i) = T(n, d-1, d-2)(1 + O(n^{-1})), \quad (41)$$

*where $n = n_i$.* □

Katona, Nemetz and Simonovits [14] showed that $T(n, k, r)/\binom{n}{r}$ is non-decreasing in $n$ and hence there exists the limit

$$t(k, r) = \lim_{n \to \infty} \frac{T(n, k, r)}{\binom{n}{r}}. \quad (42)$$

Theorem 20 and Theorem 21 essentially tell us that for fixed $d$, $T(n, d-1, d-2)$, $\rho(\mathcal{C}_i)$, and $\Gamma(n, d-2)$ are all asymptotic to $t(d-1, d-2)\binom{n}{d-2}$.[2]

**Corollary 22** *Let $\{\mathcal{C}_i\}_{i=1}^{\infty}$ be a sequence of MDS codes with strictly increasing code length $\{n_i\}_{i=1}^{\infty}$. If $d(\mathcal{C}_i) = d$ for all $i$, then*

$$\lim_{i \to \infty} \frac{\rho(\mathcal{C}_i)}{\binom{n_i}{d-2}} = \lim_{n \to \infty} \frac{\Gamma(n, d-2)}{\binom{n}{d-2}} = t(d-1, d-2). \quad (43)$$
□

The value of $t(r+1, r)$, although unknown for $r > 2$, is well-studied. In fact, the determination of $t(k, r)$ for $k > r > 2$ has been one of the most challenging open problems in combinatorial theory (for the solution of which Erdős offered a \$1000 award; see [15]). Some of the known bounds on $t(r+1, r)$ are summarized in Table I (cf. [11], [12], [8], [16], [17], [18], [19], [20]).

---

[2]Functions $f(x)$ and $g(x)$ are said to be *asymptotic to each other as $x \to x_0$* if $\lim_{x \to x_0} \frac{f(x)}{g(x)} = 1$, and is denoted by $f(x) \sim g(x)$. In this paper we usually talk about integer functions of $n$ and the condition $n \to \infty$ is sometimes omitted where there is no confusion.

TABLE I
SOME KNOWN BOUNDS ON $t(r+1, r)$

| $r$ | Lower Bound | Upper Bound |
|---|---|---|
| 2 | $\frac{1}{2}$ | $\frac{1}{2}$ |
| 3 | $\frac{9-\sqrt{17}}{12}$ | $\frac{4}{9}$ |
| 4 | $\frac{37}{143}$ | $\frac{5}{16}$ |
| 5 | $\frac{37-\sqrt{345}}{80}$ | $\frac{5}{16}$ |
| 6 | $\frac{1}{6}$ | $\frac{17}{64}$ |
| asymp. | $\frac{1}{r}$ | $\left(\frac{1}{2} + o(1)\right)\frac{\ln r}{r}$ |

In contrast, the bounds on $\rho(\mathcal{C})$ for MDS codes given in [1] are

$$\frac{1}{d-1} \leq \frac{\rho(\mathcal{C})}{\binom{n}{d-2}} \leq \frac{\max\{d^{\perp}, d-1\}}{n}. \quad (44)$$

Compared to what's promised by Corollary 22 and Table I, here the lower bound is already close to our best knowledge of $t(r+1, r)$. On the other hand, since $d^{\perp} + d - 1 = n + 1$, $\max\{d^{\perp}, d-1\}/n > 1/2$. This suggests room for improvement in the upper bound.

We will derive new upper bounds on the stopping redundancy of MDS codes through constructions of single-exclusion systems. First, consider the following construction of a Turán $(n, r+1, r)$-system due to Kim and Roush [21].

**Construction 1** Partition $[n]$ into $l$ disjoint sets, $N_0, \ldots, N_{l-1}$, with sizes as equal as possible. (For example, let $N_i := \{k \in [n] : k \equiv i \mod l\}$.) For any $X \subseteq [n]$, define

$$w(X) := \sum_{i=0}^{l-1} i|X \cap N_i|. \quad (45)$$

For $j = 0, 1, \ldots, l-1$, let

$$\begin{aligned}\mathcal{B}_j := &\{B \in [n]^r : \exists k, B \cap N_k = \emptyset\} \\ &\cup \{B \in [n]^r : w(B) \equiv j \mod l\}.\end{aligned} \quad (46)$$
□

**Theorem 23 ([21])** *For all $l$ and all $j$, $\mathcal{B}_j$ as defined in Construction 1 is a Turán $(n, r+1, r)$-system.* □

*Proof:* Let $C \in [n]^{r+1}$ be any $(r+1)$-set. If there exists $k$ such that $C \cap N_k = \emptyset$, then any $B \in [C]^r$ satisfies $B \cap N_k = \emptyset$ and hence is a member of $\mathcal{B}_j$. Otherwise, we can find $c_k \in C \cap N_k$ for all $k$. Let $B_k := C \setminus \{c_k\}$. Then $B_k \in [C]^r$. Note that $w(B_k) = w(C) - k$, $k = 0, \ldots, l-1$. So by choosing $k$ we can realize any value of $(w(B_k) \mod l)$. Therefore, for any $j$, there exists $k$ such that $B_k \in \mathcal{B}_j$. ∎

**Theorem 24** *For all $j$, $\mathcal{B}_j$ as defined in Construction 1 is a single-exclusion $(n, r)$-system if $l \geq n/(n-r-1)$.* □

*Proof:* Given Theorem 23, it suffices to show that for any $C \in [n]^i$, $i = 1, \ldots, r$, there exists $B \in \mathcal{B}_j$ such that $|C \setminus B| = 1$.

If there exists $k$ such that $C \cap N_k = \emptyset$, pick $D \in [[n] \setminus N_k]^{r+1}$ such that $C \subseteq D$. The availability of such a choice is guaranteed if $n - \lceil n/l \rceil \geq r + 1$, which is implied by $l \geq n/(n - r - 1)$. Let $B = D \setminus \{c\}$ where $c$ is an arbitrary element of $C$. Then $B \in \mathcal{B}_j$ since $B \cap N_k = \emptyset$. Also, $|C \setminus B| = |\{c\}| = 1$.

On the other hand, if for all $k$, $C \cap N_k \neq \emptyset$, we can find $c_k \in C \cap N_k$ for all $k$. Pick $D \in [n]^{r+1}$ such that $C \subseteq D$. Let $B_k := D \setminus \{c_k\}$. Similar to the proof of Theorem 23, we can show that for any $j$, there exists $k$ such that $B_k \in \mathcal{B}_j$. Also, by construction, $|C \setminus B_k| = |\{c_k\}| = 1$. ■

Now, we wish to estimate the smallest number of blocks in $\mathcal{B}_j$. Note

$$\min_{0 \leq j \leq l-1} |\mathcal{B}_j|$$
$$\leq \min_{0 \leq j \leq l-1} (|\{B \in [n]^r : \exists k, B \cap N_k = \emptyset\}|$$
$$+ |\{B \in [n]^r : w(B) \equiv j \mod l\}|) \quad (47)$$
$$= \left| \bigcup_{k=0}^{l-1} \{B \in [n]^r : B \cap N_k = \emptyset\} \right|$$
$$+ \min_{0 \leq j \leq l-1} |\{B \in [n]^r : w(B) \equiv j \mod l\}| \quad (48)$$
$$\leq \sum_{k=0}^{l-1} |\{B \in [n]^r : B \cap N_k = \emptyset\}| + \frac{1}{l}\binom{n}{r} \quad (49)$$
$$\leq l\binom{n - \lfloor \frac{n}{l} \rfloor}{r} + \frac{1}{l}\binom{n}{r}. \quad (50)$$

Therefore, we arrive at the following upper bound on $\Gamma(n, r)$.

**Theorem 25** *For all integers $l \geq n/(n - r - 1)$,*

$$\Gamma(n, r) \leq l\binom{n - \lfloor \frac{n}{l} \rfloor}{r} + \frac{1}{l}\binom{n}{r}. \quad (51)$$
□

This immediately leads to an upper bound on $\rho(\mathcal{C})$.

**Theorem 26** *Let $\mathcal{C}$ be an MDS code with length $n$ and minimum distance $d$. For all integers $l \geq R^{-1}$, where $R = (n - d + 1)/n$ is the code rate of $\mathcal{C}$,*

$$\rho(\mathcal{C}) \leq l\binom{n - \lfloor \frac{n}{l} \rfloor}{d - 2} + \frac{1}{l}\binom{n}{d - 2}. \quad (52)$$
□

Let's interpret this upper bound asymptotically as $n \to \infty$. Consider the following cases.

1) *$d$ is fixed:*
   By choosing $l = \lfloor (d-2)/(2\ln(d-2)) \rfloor$, one can see that the upper bound of Theorem 26 is better than $\frac{1 + 2\ln(d-2)}{d-2}\binom{n}{d-2}$, which is tighter than that of (44) for most values of $d$. Note that for this particular case we already knew more — Corollary 22 gives a better understanding of the asymptotic behavior of $\rho(\mathcal{C})$, and a tighter bound on $t(d-1, d-2)$ could have been used. The upper bound in Theorem 26 is valuable in that it is exact — it holds for all $n$, rather than only asymptotically in $n$.

2) *$\frac{d}{n} = \delta < 1$ is fixed:*
   Choosing $l = \lfloor (d-2)/(2\ln(d-2)) \rfloor$, we see that the upper bound of Theorem 26 is $O\left(\frac{\ln n}{n}\binom{n}{d-2}\right)$, which is better than $\Theta\left(\binom{n}{d-2}\right)$, given by (44). Note that from (44), $\rho(\mathcal{C})$ is at least $\Theta\left(\frac{1}{n}\binom{n}{d-2}\right)$.

3) *$k = n - d + 1$, the dimension of $\mathcal{C}$, is fixed:*
   Theorem 26 requires that $l \geq n/k$. If $k \geq 4$, we can choose $l$ such that $l \in (\frac{n}{3} - 1, \frac{n}{3}]$. Then the bound of Theorem 26 becomes, asymptotically,

$$\rho(\mathcal{C}) \leq l\binom{n - \lfloor \frac{n}{l} \rfloor}{d - 2} + \frac{1}{l}\binom{n}{d - 2} \quad (53)$$
$$\leq l\binom{n - 3}{n - k - 1} + \frac{1}{l}\binom{n}{n - k - 1} \quad (54)$$
$$= O(n^{k-1}) + \frac{3}{n}\left(1 + O\left(\frac{1}{n}\right)\right)\binom{n}{k+1} \quad (55)$$
$$= O(n^{k-1}) + \frac{3}{k+1}\binom{n}{k}. \quad (56)$$

The bound above is asymptotic to $\frac{3}{k+1}\binom{n}{k}$. For comparison, (44) implies an upper bound that is asymptotic to $\binom{n}{k+1}$, and a lower bound of $\frac{1}{k+1}\binom{n}{k}$.

The last case of the discussion above is interesting in its own right and we summarize it in the following theorems. Note that what we have talked about applies to $\Gamma(n, d-2) = \Gamma(n, n - k - 1)$ as well as $\rho(\mathcal{C})$.

**Theorem 27** *For fixed $k$, as $n \to \infty$,*

$$\frac{1}{k+1} \leq \frac{\Gamma(n, n - k - 1)}{\binom{n}{k}} \leq \frac{3}{k+1} + O\left(n^{-1}\right). \quad (57)$$
□

*Proof:* The lower bound is trivial since $T(n, n - k, n - k - 1) \geq \frac{1}{k+1}\binom{n}{k}$. Also, we have seen that the claimed upper bound is true for $k \geq 4$.

For $k = 3$, note that if we had been a bit more careful in writing (50), we could have shown that

$$\Gamma(n, r) \leq (l - (n \mod l))\binom{n - \lfloor \frac{n}{l} \rfloor}{r}$$
$$+ (n \mod l)\binom{n - \lfloor \frac{n}{l} \rfloor - 1}{r} + \frac{1}{l}\binom{n}{r}. \quad (58)$$

Choosing $l$ such that $l \in [\frac{n}{3}, \frac{n}{3} + 1)$ and noting $l - (n \mod l) < 3$ if $3 \nmid n$ gives the desired result.

For $k = 2$, we show that we can construct a single-exclusion $(n, n-3)$-system using less than $\frac{2}{3}\binom{n}{2}$ blocks. Let $n = 3t + r$, $r = 0, 1, 2$. Consider the $n$-set $N := ([t] \times \{0, 1, 2\}) \cup (\{t + 1\} \times \{0, \ldots, r - 1\})$. Choose as blocks the complements of the following triples (if they exist in $N$) to construct $S$:

1) $\{(x, 0), (x, 1), (x, 2)\}$, for $x = 1, \ldots, t$;
2) $\{(x, i), (y, i), (y, i + 1)\}$ and $\{(x, i), (x, i + 1), (y, i)\}$, for $x, y \in [t + 1]$, $x < y$, $i = 0, 1, 2$;
3) $\{(x, 0), (x, 2), (t + 1, 0)\}$, for $x = 1, \ldots, t$, if $r > 0$.

(In the above, $i + 1$ is modulo 3.) We claim that $S$ is a single-exclusion $(n, n - 3)$-system. Let $\iota$ be an $i$-set. We show that $\iota$ is covered in that there exists $\beta \in S$ such that $|\iota \setminus \beta| = 1$, i.e. such that $|\iota^c \cap \beta^c| = 2$. Let's call the set of points in $N$ that share a common first coordinate a *bin*. It is not hard to verify that if $\iota^c$ intersects some bin at exactly two points, then $\iota$ is covered. Also, if $\iota^c$ intersects some two bins each at just one point, then $\iota$ is also covered. Now, excluding the two cases already discussed above, we may assume that $\iota^c$ intersects no bins at two points, and intersects at most one bin at one point. But since $|\iota^c| \geq 2$, $\iota^c$ must intersect some bin at three points. This fact, however, also implies that $\iota$ is covered. Finally, it is simple algebra to verify that $|S| < \frac{2}{3}\binom{n}{2}$.

For $k = 1$, it is not hard to see that $\Gamma(n, n-2) = n - 1$. (Note in this case $T(n, n-1, n-2) = \lceil n/2 \rceil$.) ∎

The following is an immediate consequence of Theorem 27.

**Theorem 28** *Let $\{C_i\}_{i=1}^\infty$ be a sequence of MDS codes with strictly increasing code length $\{n_i\}_{i=1}^\infty$. If the dimension of $C_i$ is $k$ for all $i$, then as $i \to \infty$,*

$$\frac{1}{k+1} \leq \frac{\rho(C_i)}{\binom{n}{k}} \leq \frac{3}{k+1} + O\left(n^{-1}\right), \quad (59)$$

*where $n = n_i$.* ∎

Previously we have seen a close connection between $\Gamma(n, d-2)$ and $T(n, d-1, d-2)$. Let's see what the results of Theorem 27 and Theorem 28 tell us in those terms.

**Theorem 29** *For fixed $k$, as $n \to \infty$,*

$$\Gamma(n, n-k-1) \leq T(n, n-k, n-k-1)(3 + O(n^{-1})). \quad (60)$$

∎

*Proof:* It suffices to note that $T(n, n-k, n-k-1) \geq \frac{1}{k+1}\binom{n}{k}$, and the result follows directly from Theorem 27. It should be noted that for fixed $a$ and $b$, $T(v, v-b, v-a)$ is asymptotic to $\binom{v}{b}/\binom{a}{b}$ (cf. [22] [23]). Therefore, if $k$ is fixed, then $T(n, n-k, n-k-1) \sim \frac{1}{k+1}\binom{n}{k}$ and the claimed result is indeed the best one can get out of Theorem 27. ∎

**Theorem 30** *Let $\{C_i\}_{i=1}^\infty$ be a sequence of MDS codes with strictly increasing code length $\{n_i\}_{i=1}^\infty$. If the dimension of $C_i$ is $k$ for all $i$, then as $i \to \infty$,*

$$\rho(C_i) \leq T(n, d-1, d-2)(3 + O(n^{-1})), \quad (61)$$

*where $n = n_i$, $d = d(C_i) = n_i - k + 1$.* ∎

*Remark* The proof of Theorem 27 shows that for $k = 1, 2$,

$$\Gamma(n, n-k-1) \leq T(n, n-k, n-k-1)(2 + O(n^{-1})). \quad (62)$$

Empirical data suggest that this may be true for all $k$, so that it may be possible for the constant factor of 3 to be improved. ∎

Next, consider the following construction of a Turán $(n, r+1, r)$-system, due to Frankl and Rödl [24].

**Construction 2** Partition $[n]$ into $l$ disjoint sets, $N_0, \ldots, N_{l-1}$, with sizes as equal as possible. For all $X \subseteq [n]$, define $S(X) := \{i : X \cap N_i \neq \emptyset\}$ and $s(X) := |S(X)|$. So $s(X)$ is the number of partitions that $X$ intersects. Also, define

$$w(X) := \sum_{i=0}^{l-1} i |X \cap N_i|. \quad (63)$$

Now, for $j \in \{0, \ldots, l-1\}$, let

$$\mathcal{B}_j := \{B \in [n]^r : (w(B) + j) \mod l \\ \in \{0, 1, \ldots, l - s(B)\}\}. \quad (64)$$

∎

**Theorem 31 ([24])** *For all $l$ and all $j$, $\mathcal{B}_j$ constructed according to Construction 2 is a Turán $(n, r+1, r)$-system.* ∎

*Proof:* Note that in general, if $x \in X \cap N_i$, then $w(X \setminus \{x\}) = w(X) - i$. Let $X$ be a $(r+1)$-set. Since $X$ intersects $s(X)$ partitions, $\{(w(Y) + j) \mod l : Y \in [X]^r\}$ contains $s(X)$ distinct values. Hence, there exists $Y \in [X]^r$, such that $(w(Y) + j) \mod l \in \{0, 1, \ldots, l - s(X)\}$. Now, note that $s(Y) \leq s(X)$ since $Y \subseteq X$. Therefore, $(w(Y) + j) \mod l \in \{0, 1, \ldots, l - s(Y)\}$, which implies that $Y \in \mathcal{B}_j$. ∎

**Theorem 32** *If $n \geq l(r+1)$, then for all $j$, $\mathcal{B}_j$ constructed according to Construction 2 is a single-exclusion $(n, r)$-system.* ∎

*Proof:* Given Theorem 31, it suffices to show that all $i$-sets are covered by $\mathcal{B}_j$, $i = 1, \ldots, r$.

Let $X$ be an $i$-set. Choose $Z \in [n]^{(r+1)}$, such that $X \subseteq Z$ and $S(Z) = S(X)$. This is possible as $|\bigcup_{k \in S(X)} N_k| \geq s(X)(r+1) \geq r+1$. Consider the class of $r$-sets, $\mathcal{Y} := \{Z \setminus \{x\} : x \in X\}$. Note that $\{(w(Y) + j) \mod l : Y \in \mathcal{Y}\}$ contains $s(X)$ distinct values. Hence, there exists $Y \in \mathcal{Y}$, such that $(w(Y) + j) \mod l \in \{0, 1, \ldots, l - s(X)\}$. Now, note that $Y \subseteq Z$ implies that $s(Y) \leq s(Z) = s(X)$. Therefore, $(w(Y) + j) \mod l \in \{0, 1, \ldots, l - s(Y)\}$, which implies that $Y \in \mathcal{B}_j$. Finally, it is clear that $|X \setminus Y| = 1$. ∎

Now we wish to estimate $\min_j |\mathcal{B}_j|$. It can be shown that ([8])

$$\sum_{j=0}^{l-1} |\mathcal{B}_j| = \binom{n}{r} + l \binom{n - \lfloor \frac{n}{l} \rfloor}{r}. \quad (65)$$

Therefore,

$$\min_j |\mathcal{B}_j| \leq \frac{1}{l} \sum_{j=0}^{l-1} |\mathcal{B}_j| = \frac{1}{l}\binom{n}{r} + \binom{n - \lfloor \frac{n}{l} \rfloor}{r}. \quad (66)$$

Thus, we have the following theorems.

**Theorem 33** *For all integers $l \leq n/(r+1)$,*

$$\Gamma(n, r) \leq \frac{1}{l}\binom{n}{r} + \binom{n - \lfloor \frac{n}{l} \rfloor}{r}. \quad (67)$$

∎

**Theorem 34** *Let $\mathcal{C}$ be an MDS code with length $n$ and minimum distance $d$. Then for all integers $l \leq (1-R)^{-1}$, where $R = (n-d+1)/n$ is the code rate of $\mathcal{C}$,*

$$\rho(\mathcal{C}) \leq \frac{1}{l}\binom{n}{d-2} + \binom{n - \lfloor \frac{n}{l} \rfloor}{d-2}. \quad (68)$$

∎

The requirement that $l$ be no greater than $(1-R)^{-1}$ turns out to be too restrictive for most cases and makes the upper bound less useful when $R$ is not close to 1. To mitigate the problem, we can get rid of this requirement by adding some more blocks to $\mathcal{B}_j$. For the sake of discussion, let's first assume $l \mid n$.

**Construction 3** Arrange elements of $[n]$ into a $\frac{n}{l} \times l$ matrix (in an arbitrary way). The columns of this matrix partition $[n]$ into $l$ disjoint sets with equal size which we denote by $N_0, \ldots, N_{l-1}$. With $N_0, \ldots, N_{l-1}$, let $\mathcal{B}_j$ be defined the same way as described in Construction 2. Now, the rows of this matrix also partition $[n]$. We denote them by $M_0, \ldots, M_{\frac{n}{l}-1}$. For all $X \subseteq [n]$, define

$$w'(X) := \sum_{i=0}^{\frac{n}{l}-1} i |X \cap M_i|. \quad (69)$$

For $t = 0, \ldots, \frac{n}{l} - 1$, let

$$\mathcal{M}_t := \left\{ B \in [n]^r : w'(B) \equiv t \mod \frac{n}{l} \right\}. \quad (70)$$

Finally, for all $j$, $t$, let

$$\mathcal{B}_{j,t} := \mathcal{B}_j \cup \mathcal{M}_t. \quad (71)$$

We show that $\mathcal{B}_{j,t}$ as defined in Construction 3 is a single-exclusion $(n,r)$-system for all $l$.

**Lemma 35** *Let $l \geq 2$ be an integer. Let $L = \{0, 1, \ldots, l-1\}$. For all $X \subseteq L$, define*

$$\|X\| := \sum_{i \in X} i. \quad (72)$$

*Then, for all $k$, $k = 1, \ldots, l-1$,*

$$\{\|\kappa\| \mod l : \kappa \in [L]^k\} = L. \quad (73)$$

*Proof:* First, it is easy to see that the claim is true for $k = 1$ and $2$. The case $k = 1$ is quite trivial. For $k = 2$, it suffices to note that $i = \|\{0, i\}\|$ for $i = 1, \ldots, l-1$, and $0 = \|\{1, l-1\}\|$.

In general, if the claim is true for $k = m$, then it is also true for $k = l - m$, since

$$\{\|\kappa\| \mod l : \kappa \in [L]^{l-m}\}$$
$$= \{\|L\| - \|\kappa\| \mod l : \kappa \in [L]^m\}. \quad (74)$$

So, the claim is also true for $k = l - 1$ and $k = l - 2$.

Now, for the general case, let's assume $k \leq l - 3$. The idea is to consider pairs of elements in $L$ that sum to 0 modulo $l$. First, suppose $l$ is even. Then $L$ can be partitioned in the following way:

$$L = \{0\} \cup \{l/2\} \cup \bigcup_{i=1}^{l/2-1} \{i, l-i\} = \{0\} \cup \{l/2\} \cup \bigcup_{i=1}^{l/2-1} Z_i, \quad (75)$$

where $Z_i := \{i, l-i\}$, $i = 1, \ldots, l/2 - 1$. We show that for all $j \in L$, we can find a $k$-set $\beta$ such that $\|\beta\| \equiv j \mod l$. If $k$ is even, then:

- If $j \in Z_m$ for some $m$, let $\beta$ be the union of $\{0, j\}$ and $(k/2 - 1)$ $Z_i$'s other than $Z_m$.
- If $j = l/2$, let $\beta$ be the union of $\{0, j\}$ and $(k/2 - 1)$ $Z_i$'s.
- If $j = 0$, let $\beta$ be the union of $k/2$ $Z_i$'s.

Similarly, if $k$ is odd, then:

- If $j \in Z_m$ for some $m$, let $\beta$ be the union of $\{j\}$ and $(k-1)/2$ $Z_i$'s other than $Z_m$.
- If $j = l/2$, let $\beta$ be the union of $\{j\}$ and $(k-1)/2$ $Z_i$'s.
- If $j = 0$, let $\beta$ be the union of $\{0\}$ and $(k-1)/2$ $Z_i$'s.

For odd $l$, the proof is very similar and we will not elaborate here. ∎

**Theorem 36** *For all $l$, $j$, and $t$, $\mathcal{B}_{j,t}$ as defined in Construction 3 is a single-exclusion $(n,r)$-system.* □

*Proof:* Let $X$ be an $i$-set, $i = 1, \ldots, r$, and $x \in X$ be an arbitrary element. First, suppose that for all $k$, $N_k \not\subseteq X$. If $r \leq n - l$, then we can find an $r$-set $Z \supseteq X$ such that $N_k \not\subseteq Z$ for all $k$. Now, choose $y_k \in N_k \setminus Z$ for all $k$ and consider $r$-sets of the form $Y_k := (Z \setminus \{x\}) \cup \{y_k\}$. For all $j$, we can choose $k$ such that $w(Y_k) + j \equiv 0 \mod l$, and hence $Y_k \in \mathcal{B}_j$. Clearly, $|X \setminus Y_k| = 1$. On the other hand, if $r > n - l$, then we can find an $(n-l)$-set $Z \supseteq X$ such that $N_k \not\subseteq Z$ for all $k$. Clearly, $[n] \setminus Z$ intersects each $N_k$ at exactly one element. Consider $r$-sets that consist of the union of $Z \setminus \{x\}$ and an $(r-n+l+1)$-subset of $[n] \setminus Z$. By Lemma 35, for all $j$, there exists $W \in [[n] \setminus Z]^{r-n+l+1}$ such that if $Y = (Z \setminus \{x\}) \cup W$ then $w(Y) + j \equiv 0 \mod l$. Therefore, $Y \in \mathcal{B}_j$ and clearly $|X \setminus Y| = 1$.

Otherwise, suppose $N_k \subseteq X$. By construction, $N_k$ contains elements from each $M_m$. Let $Z \supseteq X$ be an $(r+1)$-set; then, by choosing $Y \in \mathcal{Y} := \{Z \setminus \{x\} : x \in X\}$, we can realize any value of $w'(Y)$. Hence, for any $t$, there exists an $r$-set $Y \in \mathcal{M}_t$ such that $|X \setminus Y| = 1$. ∎

If $l \nmid n$, we can define $M_0, \ldots, M_{\lfloor \frac{n}{l} \rfloor - 1}$ by applying Construction 3 to the first $\lfloor n/l \rfloor l$ elements of $[n]$ and letting $M_{\lfloor \frac{n}{l} \rfloor - 1}$ include the extra $(n \mod l)$ elements. All reasoning is still valid.

Clearly,

$$\sum_{t=0}^{\lfloor \frac{n}{l} \rfloor - 1} |\mathcal{M}_t| = \binom{n}{r}. \quad (76)$$

Hence,

$$\min_t |\mathcal{M}_t| \leq \frac{1}{\lfloor n/l \rfloor} \binom{n}{r}. \quad (77)$$

By the union bound, $|\mathcal{B}_{j,t}| \leq |\mathcal{B}_j| + |\mathcal{M}_t|$, hence we arrive at the following bounds.

**Theorem 37** *For all integers $l$,*

$$\Gamma(n,r) \leq \begin{cases} \binom{n-\lfloor n/l \rfloor}{r} + \frac{1}{l}\binom{n}{r} & \text{if } l \leq \frac{n}{r+1} \\ \binom{n-\lfloor n/l \rfloor}{r} + \left(\frac{1}{l} + \frac{1}{\lfloor n/l \rfloor}\right)\binom{n}{r} & \text{if } l > \frac{n}{r+1}. \end{cases} \quad (78)$$
□

**Theorem 38** *Let $\mathcal{C}$ be an MDS code with length $n$ and minimum distance $d$. Then for all integers $l$,*

$$\rho(\mathcal{C}) \leq \begin{cases} \binom{n-\lfloor n/l \rfloor}{d-2} + \frac{1}{l}\binom{n}{d-2} & \text{if } l \leq (1-R)^{-1} \\ \binom{n-\lfloor n/l \rfloor}{d-2} + \left(\frac{1}{l} + \frac{1}{\lfloor n/l \rfloor}\right)\binom{n}{d-2} & \text{if } l > (1-R)^{-1} \end{cases}, \quad (79)$$

*where $R = (n-d+1)/n$ is the code rate of $\mathcal{C}$.* □

Note that when we choose $l$ in the region $l > (1-R)^{-1}$, the upper bound is never better than $\frac{2}{\sqrt{n}}\binom{n}{d-2}$. So the strength of the bound above still lies in the regime of high rate codes.

Figures 1 through 3 compare the upper bounds we have obtained so far, i.e. those of Theorem 26 and Theorem 38, to the previously known bounds of (44). In the plots, all bounds are normalized with respect to $\binom{n}{d-2}$. We see that both newly proposed upper bounds are tighter than (44) in a variety of situations, with the one based on Construction 1 outperforming the one based on Construction 3 for all but very high code rate scenarios.

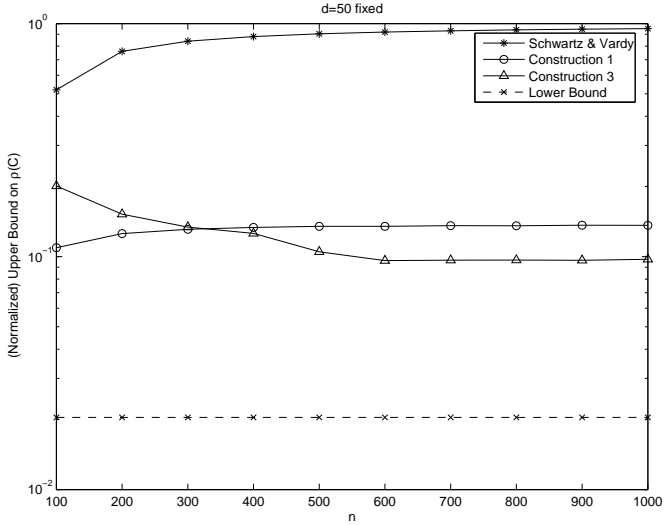

Fig. 1. Bounds on $\rho(\mathcal{C})$ for $(n,k,d)$ MDS codes. $d=50$ is fixed. Bounds are normalized relative to $\binom{n}{d-2}$.

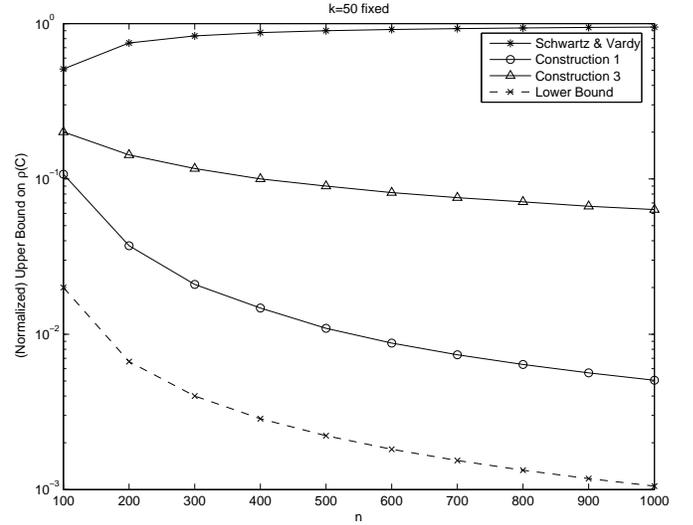

Fig. 3. Bounds on $\rho(\mathcal{C})$ for $(n,k,d)$ MDS codes. $k=50$ is fixed. Bounds are normalized relative to $\binom{n}{d-2}$.

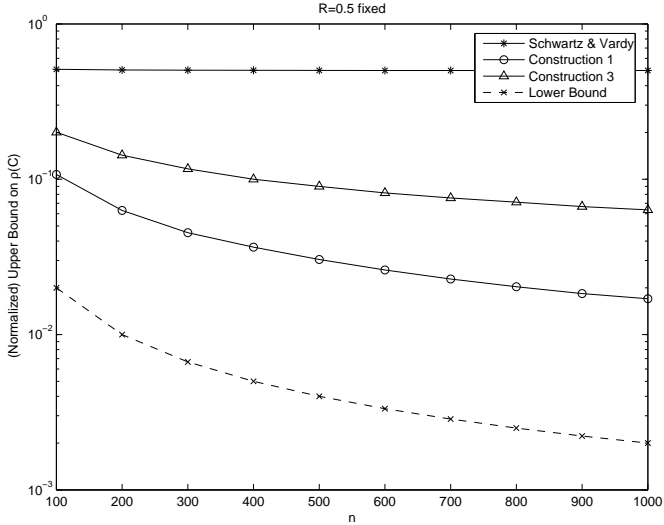

Fig. 2. Bounds on $\rho(\mathcal{C})$ for $(n,k,d)$ MDS codes. $R=0.5$ is fixed. Bounds are normalized relative to $\binom{n}{d-2}$.

## IV. CONCLUSION

We have obtained new upper bounds on the stopping redundancy of linear codes. Compared to the previously known bounds of [1], [2], our bound based on the "probabilistic method" gives better results for a number of interesting cases, including for all "good" codes, i.e. those whose minimum distance is asymptotically non-trivial relative to code length.

Though tighter, the new upper bounds for the case of "good" codes are still exponential in the length of the code. It remains an open question whether there exist "good" codes whose stopping redundancy is polynomial in the code length.

Improving the lower bound on stopping redundancy seems to be difficult. Applying the probabilistic method only yields the same bound as given in [1].

For MDS codes, the interesting relationship between stopping redundancy and Turán numbers has been explored. We have defined a new combinatorial quantity, the single-exclusion number $\Gamma(v,r)$, and related it to the Turán number and the stopping redundancy of MDS codes. By studying $\Gamma(v,r)$, we have obtained new upper bounds on the stopping redundancy of MDS codes, which have been shown to be tighter than the best previously known bounds for various situations. We have also proved that for MDS codes with length $n$ and minimum distance $d$, $\rho(\mathcal{C})$ is asymptotic to $T(n,d-1,d-2)$ for fixed $d$, and is asymptotic to $T(n,d-1,d-2)$ up to a constant factor of at most 3 for fixed $k=n-d+1$. We conjecture that in the latter case the constant factor can be improved to 2. We also conjecture that $\rho(\mathcal{C}) = \Gamma(n,d-2)$ for all MDS codes. For one thing, the two are asymptotic to each other if $d$ is fixed. Further, for $d=3,4$, both $\rho(\mathcal{C})$ and $\Gamma(n,d-2)$ are equal to $T(n,d-1,d-2)$. For $d=5$, we have shown that neither can differ from $T(n,d-1,d-2)$ by more than 1.

## APPENDIX I
## THE BINARY GOLAY CODE

We present here a parity-check matrix with 34 rows that achieves maximum stopping distance and corrects more low-weight erasure patterns than the parity-check matrix given in [2]. The details of our parity-check matrix, denoted by $H$, are given in Table II. It was found by a greedy computer search. The idea is to start with a random selection of codewords from $\mathcal{G}_{24}$ (note that $\mathcal{G}_{24}$ is self-dual), and in each iteration, replace one codeword in the selection so that as many more $i$-sets as possible are covered. When no such improvements can be

TABLE II
PARITY CHECK MATRIX WITH 34 ROWS FOR $\mathcal{G}_{24}$ THAT ACHIEVES STOPPING DISTANCE 8

$$H = \begin{pmatrix}
0 0 0 0 0 0 0 1 1 0 1 1 0 1 0 0 0 0 1 1 1 0 0 0 \\
0 0 0 0 0 0 1 0 0 1 0 0 0 1 1 1 1 0 1 0 0 0 0 1 \\
0 0 0 0 0 0 1 1 1 0 1 0 0 0 0 1 0 1 0 0 1 0 1 0 \\
1 0 0 0 0 1 0 0 1 0 0 1 0 0 1 0 1 0 0 0 0 1 1 0 \\
0 0 0 0 0 1 0 0 1 1 1 0 1 1 0 0 0 0 1 0 0 0 0 1 \\
0 0 0 0 0 1 1 0 0 0 0 1 1 1 1 0 0 0 1 0 0 1 1 0 \\
0 0 0 0 0 1 1 1 0 0 1 1 1 0 0 0 0 0 0 1 0 0 1 \\
1 0 0 0 1 0 0 0 0 0 0 0 1 0 1 1 0 1 1 1 0 0 \\
1 0 0 0 1 0 0 1 0 1 1 1 1 0 0 0 0 0 1 0 0 0 0 0 \\
0 0 0 0 1 0 1 0 0 0 0 0 0 1 0 0 1 1 0 1 0 1 1 \\
1 0 0 0 1 1 1 0 0 0 0 0 0 1 1 0 0 0 0 0 1 0 1 \\
0 0 0 1 0 0 0 1 0 0 1 1 1 0 0 0 0 1 0 1 0 0 1 0 \\
1 0 0 1 0 0 0 0 1 0 0 0 0 1 0 0 0 1 1 0 0 1 0 1 0 \\
0 0 0 1 0 1 0 1 0 0 0 1 0 1 0 1 0 1 0 0 1 0 0 0 \\
1 0 0 1 1 0 0 1 1 1 0 0 0 0 1 0 0 0 0 0 0 1 0 0 \\
0 0 1 0 0 0 0 0 1 0 0 0 0 1 0 1 0 1 0 1 1 1 0 0 \\
0 0 1 0 0 0 1 1 1 0 1 1 1 0 0 0 0 0 0 0 0 0 1 \\
0 0 1 0 0 1 1 0 0 1 0 0 1 0 1 0 0 1 0 0 0 0 0 1 \\
0 0 1 0 1 1 0 0 0 0 0 1 0 1 0 1 0 0 0 0 0 1 1 0 \\
0 0 1 1 0 0 0 0 1 0 1 0 1 1 1 0 0 1 0 1 0 0 0 0 \\
0 0 1 1 1 1 0 0 0 0 1 0 1 0 1 1 0 0 1 0 0 0 0 1 \\
0 1 0 0 0 0 0 0 1 1 0 1 0 0 1 0 0 0 1 1 1 0 0 \\
1 1 0 0 0 0 1 0 0 1 0 0 0 0 0 1 0 1 0 0 0 0 1 1 \\
0 1 0 0 0 1 0 0 0 1 1 0 0 0 1 1 0 0 0 1 0 0 0 1 \\
0 1 0 0 1 0 0 1 0 0 0 1 0 1 0 1 1 0 1 0 0 0 0 0 \\
0 1 0 0 1 1 0 0 0 0 0 1 0 0 0 0 1 0 0 1 0 1 0 1 \\
1 1 0 1 0 0 1 1 0 1 0 0 0 0 0 0 0 1 0 1 0 0 0 0 \\
0 1 0 1 0 1 0 1 0 0 0 0 0 1 0 0 1 1 0 0 1 0 0 0 \\
0 1 0 1 1 0 0 1 0 0 0 0 1 0 1 0 0 0 1 0 0 0 1 0 \\
1 1 1 0 0 0 0 0 0 0 0 1 0 0 0 0 0 1 0 1 0 1 1 0 \\
1 1 1 0 0 0 1 0 1 0 1 0 0 0 0 0 1 0 0 0 0 0 1 0 \\
1 1 1 0 1 0 0 0 0 0 0 0 1 0 0 0 1 0 1 1 0 0 0 0 \\
1 1 1 1 0 0 0 0 1 1 0 0 0 0 0 1 0 0 0 1 0 0 0 0 \\
0 1 1 1 1 0 0 0 0 0 0 0 1 1 0 1 0 0 0 0 1 0 0 0
\end{pmatrix}$$

TABLE III
NUMBER OF UNDECODABLE ERASURE PATTERNS BY WEIGHT $w$ FOR DIFFERENT ITERATIVE DECODERS FOR $\mathcal{G}_{24}$

| $w$ | $\Psi_H(w)$ | $\Psi_{H'_{24}}(w)$ | $\Psi_{\mathrm{ML}}(w)$ |
|---|---|---|---|
| $\leq 7$ | 0 | 0 | 0 |
| 8 | 3284 | 3598 | 759 |
| 9 | 78218 | 82138 | 12144 |
| 10 | 580166 | 585157 | 91080 |
| 11 | 1734967 | 1717082 | 425040 |
| 12 | 2569618 | 2556402 | 1313116 |
| $\geq 13$ | $\binom{24}{w}$ | $\binom{24}{w}$ | $\binom{24}{w}$ |

made, an additional codeword is added to the selection and the iteration continues. The process is stopped when the desired stopping distance is achieved. We find that it is enough to only consider covering 7-sets, and verify in the end that the matrix obtained indeed covers all smaller $i$-sets and has the proper rank.

Table III compares the number of undecodable erasure patterns by weight $w$ (number of erased bits) for iterative decoders based on $H$, $H'_{24}$ (the 34-row parity-check matrix reported in [2]), and the maximum-likelihood decoder. We see that the iterative decoder based on $H$ corrects considerably more lower weight erasure patterns than does the one based on $H'_{24}$, which implies that it will perform better when the erasure probability is small. For a binary erasure channel with erasure probability $p$, a detailed comparsion shows that for all $p < 0.349$, the iterative decoder based on $H$ has a smaller probability of decoding failure.

## APPENDIX II
## DERIVATIONS IN THE ASYMPTOTIC COMPARISON OF BOUNDS

### A. Binary Linear Codes (Example 2, Case 1)

Noting that $-\log(1-x) \sim \frac{x}{\ln 2}$ as $x \to 0$, we see the upper bound in (16) is $O(2^{\delta n})$, hence so is the bound in Theorem 3.

On the other hand, note that

$$\sum_{i=1}^{d(\mathcal{C})-1} \binom{n}{i}\left(1 - \frac{i}{2^i}\right)^\rho \geq \binom{n}{d(\mathcal{C})-1}\left(1 - \frac{d(\mathcal{C})-1}{2^{d(\mathcal{C})-1}}\right)^\rho. \tag{80}$$

Setting

$$\binom{n}{d(\mathcal{C})-1}\left(1 - \frac{d(\mathcal{C})-1}{2^{d(\mathcal{C})-1}}\right)^\rho = 1, \tag{81}$$

and solving for $\rho$, one can readily show that $\rho^*(n, d(\mathcal{C}))$ is also $\Omega(2^{\delta n})$. Therefore, the bound given by Theorem 3 is indeed $\Theta(2^{\delta n})$.

In comparison, consider the bound in Theorem 1. For $0 < \delta < \frac{1}{2}$, the asymptotic Plotkin bound implies that $\frac{\delta}{\gamma} \leq \frac{1}{2}$. Noting that $h(p) \geq 2p$ for $p \leq \frac{1}{2}$, we have

$$\sum_{i=1}^{d(\mathcal{C})-2} \binom{r(\mathcal{C})}{i} = \Omega\left(\binom{r(\mathcal{C})}{d(\mathcal{C})-2}\right) \tag{82}$$

$$= \Omega\left(\binom{\gamma n}{\delta n - 2}\right) \tag{83}$$

$$= \Omega\left(\binom{\gamma n}{\delta n}\right) \tag{84}$$

$$= \Omega\left(\frac{1}{\sqrt{n}} 2^{\gamma h(\frac{\delta}{\gamma})n}\right) \tag{85}$$

$$= \Omega\left(\frac{2^{2\delta n}}{\sqrt{n}}\right). \tag{86}$$

The analysis for the bound of Theorem 2 is similar, and one can show that the same asymptotic result applies.

### B. Linear Codes over $\mathbb{F}_q$ (Example 4)

Showing that the bound in Theorem 7 is $\Theta(q^{\delta n})$ is very similar to the binary case, and we will not elaborate here.

Now, consider the bound of Theorem 6. Let $\theta = \frac{q-1}{q}$. For $0 < \delta < \theta$, we see that $0 < \frac{\delta}{\gamma} \leq \theta$ by the asymptotic Plotkin bound. Noting that for all $0 < \theta < 1$, $h(p) \geq \frac{h(\theta)}{\theta} p$ for

$0 < p \leq \theta$, we have

$$\sum_{i=1}^{d(\mathcal{C})-1} \binom{r(\mathcal{C})}{i}(q-1)^{i-1}$$

$$= \Omega\left(\binom{r(\mathcal{C})}{d(\mathcal{C})-1}(q-1)^{d(\mathcal{C})-2}\right) \quad (87)$$

$$= \Omega\left(\frac{1}{\sqrt{n}}2^{\gamma h(\frac{\delta}{\gamma})n}(q-1)^{\delta n}\right) \quad (88)$$

$$= \Omega\left(\frac{1}{\sqrt{n}}2^{\gamma \frac{h(\theta)}{\theta}\frac{\delta}{\gamma}n}(q-1)^{\delta n}\right) \quad (89)$$

$$= \Omega\left(\frac{1}{\sqrt{n}}\left(\frac{1}{\theta}\right)^{\delta n}\left(\frac{1}{1-\theta}\right)^{\frac{1-\theta}{\theta}\delta n}(q-1)^{\delta n}\right) \quad (90)$$

$$= \Omega\left(\frac{1}{\sqrt{n}}q^{\frac{q}{q-1}\delta n}\right). \quad (91)$$

## ACKNOWLEDGMENT

The authors would like to thank Moshe Schwartz for helpful discussions.